%% file: apssamp.tex
\documentclass[aps,prl,amsmath, amssymb,twocolumn,superscriptaddress,nofootinbib]{revtex4-2}

\usepackage{graphicx}% Include figure files
\usepackage{dcolumn}% Align table columns on decimal point
\usepackage{bm}% bold math
\usepackage{aasmacros}
\usepackage{hyperref}% add hypertext capabilities
\usepackage{xcolor}

\usepackage{soul}
\usepackage[normalem]{ulem}
\newcommand{\mpcph}{~h^{-1}{\rm Mpc}}
\newcommand{\msolph}{~h^{-1}{\rm M}_{\odot}}

\newcommand{\RR}{$\mathcal{R}$}
\newcommand{\LL}{$\mathcal{L}$}

\definecolor{bubbles}{rgb}{0.91, 1.0, 1.0}
\definecolor{aquamarine}{rgb}{0.5, 1.0, 0.83}
\definecolor{bubblegum}{rgb}{0.99, 0.76, 0.8}
\definecolor{bluebell}{rgb}{0.84, 0.84, 0.99}
\definecolor{dollarbill}{rgb}{0.72, 0.93, 0.6}

\newcounter{UPcounter}

\newcounter{HYcounter}

\newcounter{TOcounter}
    
\newcounter{JScounter}

\begin{document}

\preprint{APS/123-QED}

\title{Probing vector chirality in the early Universe}% Force line breaks with \\

\author{Junsup Shim}\email{jshim@asiaa.sinica.edu.tw}
\affiliation{Academia Sinica Institute of Astronomy and Astrophysics (ASIAA), No. 1, Section 4, Roosevelt Road, Taipei 10617, Taiwan}
\author{Ue-Li Pen}
\affiliation{Academia Sinica Institute of Astronomy and Astrophysics (ASIAA), No. 1, Section 4, Roosevelt Road, Taipei 10617, Taiwan}
\affiliation{Canadian Institute for Theoretical Astrophysics, University of Toronto, 60 St. George Street, Toronto, ON M5S 3H8, Canada}
\affiliation{Department of Physics, University of Toronto, 60 St. George Street, Toronto, ON M5S 1A7, Canada}
\affiliation{Dunlap Institute for Astronomy \& Astrophysics, University of Toronto, 50 St. George Street, Toronto, ON M5S 3H4, Canada}
\affiliation{Perimeter Institute for Theoretical Physics, 31 Caroline Street North, Waterloo, ON N2L 2Y5, Canada}
\affiliation{Canadian Institute for Advanced Research, CIFAR Program in Gravitation and Cosmology, 661 University Avenue, Toronto, ON M5G 1M1, Canada}
\author{Hao-Ran Yu}
\affiliation{Department of Astronomy, Xiamen University, Xiamen, Fujian 361005, China}
\author{Teppei Okumura}
\affiliation{Academia Sinica Institute of Astronomy and Astrophysics (ASIAA), No. 1, Section 4, Roosevelt Road, Taipei 10617, Taiwan}
\affiliation{Kavli Institute for the Physics and Mathematics of the Universe (WPI), UTIAS, The University of Tokyo, Kashiwa, Chiba 277-8583, Japan}

\date{\today}% It is always \today, today,

\begin{abstract}
We explore the potential of using late-time galaxy spins to test the parity symmetry of primordial vector fossils. Using $N$-body simulations, we analyze halo spins as a reliable proxy for galaxy spins to investigate the detectability of this effect.
We develop a novel approach to generate initial conditions (ICs) that have substantial parity asymmetry but do not alter the initial matter power spectrum.
We construct the initial spin fields from the parity broken ICs and halo spin fields using late-time halos evolved from such ICs.
Focusing on the helicity of these vector fields, we detect substantial asymmetry in the initial spin field. In addition, we find that over $50\%$ of the initial spin field's asymmetry remains in the late-time halo spin field on a range of scales.
Based on mock galaxy spin fields derived from the halo spin fields, we forecast that a maximum detection at $13\sigma$ is possible with the final DESI BGS for the model considered in this analysis.
Our findings demonstrate that primordial vectorial parity violation survives nonlinear gravitational
evolution, and thus, can be effectively probed with galaxy spins at late times.

\end{abstract}

\maketitle

Recently, unexpected $3-7\sigma$ signals of parity asymmetry were detected \cite{philcox22,hou+23} in observation using parity-odd basis \cite{cahn+23,cahn&slepian23}, with a focus on the ``\textit{helicity}'' (or handedness) of galaxy quartets in 3 dimensions.
Although particular care must be taken for accurately modeling systematic uncertainties to conclude such a signal to be of primordial origin \cite{hou+22,hou+23,cabass+23,philcox23,philcox&ereza24,adari&slosar24}, this motivates further exploration for various types of potential parity violation (PV) in the early Universe and their signatures.

The primary focus has been on the theoretical search for tensorial type, where some inflationary models can produce unequal primordial gravity waves (pGWs) with left- and right-helicities \cite{lue+99, jackiw&pi03, satoh+08, sorbo11, maleknejad+13,bartolo&orlando17}. Thus, detecting either the helical asymmetry between two chiral-pGWs or their imprints on CMB polarization \cite{saito+07, contaldi+08, gluscevic&kamionkowski10,eiichiro22} and LSS \cite{jeong&kamionkowski10, masui+17, biagetti&orlando20, philcox+24shape, okumura&sasaki24} could confirm the primordial tensorial-PV.
Such helical asymmetry imprints would be more detectable in 3-dimensional probes \cite{masui+17} since the helicity information is largely lost in 2 dimensions, for example, leaving the detection in the polarized CMB \cite{gluscevic&kamionkowski10, gerbino+16,philcox&shiraishi24, philcox&shiraishi24grav} more challenging \cite[c.f. see][hinting at the non-vanishing EB-correlation of non-primordial origin, e.g., cosmic birefringence]{minami&komatsu20, eskilt&komatsu22,diego-palazuelos+22}.

Another interesting, but largely unexplored, class of parity violation is vectorial type, possibly generating helically asymmetric vector fossils \cite{jeong&kamionkowski10} that may leave imprints in initial vector quantities.
Recently, a 3-dimensional estimator ``\textit{initial spin}'' (rigorously defined in Eq.~\ref{eqn:initspin}) -- a vector representing the spinning of a mass element torqued by the initial tidal field \cite{peebles69, doroshkevich70, white84} -- was proposed \cite{yu+20,motloch+21} and examined \cite{motloch+22} to search for vectorial helical asymmetry.

This proposal relies on two key properties of the initial spin field --- 1) a vector field that only consists of left- and right-helical modes, and 2) its strong alignments with ``\textit{galaxy spin}'' and ``\textit{halo spin}'', or the angular momentum direction of a galaxy \cite{motloch+21,sheng+23} and a dark matter halo \cite{lee&pen00, porciani+02,jiang+19,yu+20} at late time.
The latter property suggests that the characteristics of the initial spin field will remain in the galaxy spin field.
Therefore, detecting helical asymmetry in a galaxy spin field would indicate that the initial spin field is helically asymmetric, evidencing vectorial-PV in the early Universe.

On the observational side, there has not yet been compelling evidence that the galaxy spin field is helically asymmetric. It has only been shown that there are mildly uneven correlations to the left- and right-helical components of the reconstructed initial spin field \cite{motloch+22}.
However, it is not straightforward to conclude that primordial PV is absent due to such non-detection. This is because galaxies' nonlinear formation and evolution may have substantially diminished the primordial asymmetry, leaving only a weak or near-parity symmetric signal in galaxy spins.
Thus, it is necessary to further investigate into how effectively the primordial helical asymmetry in the initial spin is carried onto the late-time galaxy spin.

In this Letter, we investigate the detectability of PV in a primordial vector fossil using galaxy spins. Specifically, we examine the PV in a ``\textit{helical}'' (i.e., rotational) displacement vector field.
The helical displacement vector field is envisioned to be produced in the very early universe, e.g., during inflation, and remains as fossils \cite{jeong&kamionkowski10}.
Note that the initial vector mode perturbations decay over time but can leave vector fossils\footnote{We confirmed that the helical displacement remains a fossil in the linear regime, as its effect on the spin field at a fixed time is consistent across simulations initialized at different redshifts, all with the same fixed amplitude of the helical displacement.}, in analogy to tensor perturbations imprinting tensor fossils \cite{masui&pen10, schmidt+14, masui+17}.

With the help of numerical N-body simulations, we analyze halo spins as a reliable proxy for galaxy spins, leveraging the tight correlation between those two spins \cite{hahn+10, teklu+15, jiang+19, sheng+23}.
In the spirit of effective field theory, without prior assumptions on microscopic physics, we develop a novel way of generating the initial condition with a helically asymmetric initial spin field.
We show that the parity-broken displacement field efficiently produces strong primordial helical asymmetry without affecting the initial matter power spectrum. Our analysis of halos evolved from these initial conditions reveals that this asymmetry survives significantly in the final halo spin field.

\textit{Simulations and initial conditions.}---
We construct 500 pairs of N-body simulations -- ones with parity symmetric initial condition and the others with parity asymmetric initial condition.
The simulations adopt the modified version of the code {\small CUBE} \cite{yu+18} to evolve $256^{3}$ dark matter particles within a periodic cubic box of size $L=100\mpcph$. For numerical convenience, the initial redshift is set $z_{\rm init}=30$.
We generate the initial density fluctuation of the standard parity symmetric universe using Zel'dovich approximation \cite{zeldovich70} where the displacement vector field, $\bm{\Psi}_{\rm ZA}$, is curl-free (i.e., irrotational) and its Fourier space representation follows
\begin{equation}\label{eqn:dspZA}
    % \bm{\Psi}_{\rm sy}=\bm{\Psi}_{\rm ZA}=-i\bm{k}\frac{\Phi(\bm{k})}{4\pi G a^{2}\bar{\rho}}\ ,
    \bm{\Psi}_{\rm ZA}=-i\bm{k}\frac{\Phi(\bm{k})}{4\pi G a^{2}\bar{\rho}}\ ,
\end{equation}
with the gravitational potential $\Phi$, scale factor $a$, gravitational constant $G$, and mean cosmic density $\bar{\rho}$.
On the other hand, the parity asymmetric initial condition is constructed using the modified displacement field, $\bm{\Psi}_{\rm tot}$, which includes an extra helical displacement, $\bm{\Psi}_{\rm hel}$,
\begin{equation}
     % \bm{\Psi}_{\rm asy} = \bm{\Psi}_{\rm ZA}+\bm{\Psi}_{\rm hel}.
     \bm{\Psi}_{\rm tot} = \bm{\Psi}_{\rm ZA}+\bm{\Psi}_{\rm hel}.
\end{equation}

The helical displacement field can contain two independent components that are purely right (\RR)- and left (\LL)-helical,
\begin{equation}
    \begin{aligned}
     \bm{\Psi}_{\rm hel} &= \bm{\Psi}_{\mathcal{R}}+\bm{\Psi}_{\mathcal{L}}=\sum_{\mathcal{H}=\mathcal{R},\mathcal{L}}\Phi_{\mathcal{H}}(\bm{k})\hat{\bm{e}}_{\mathcal{H}}(\bm{k}).
     \end{aligned}
\end{equation}
The \RR- and \LL-helical displacement fields can be constructed via $\bm{\Psi}_{\mathcal{R}/\mathcal{L}}=\Phi_{\mathcal{R/L}}\hat{\bm{e}}_{\mathcal{R/L}}$, with a pseudo-scalar field, $\Phi_{\mathcal{R}/\mathcal{L}}\equiv\hat{\bm{e}}_{\mathcal{R/L}}\bm{\Psi}_{\rm tot}$, and an eigenvector, 
\begin{equation}
    \hat{\bm{e}}_{\mathcal{R}/\mathcal{L}}(\bm{k}) =\frac{1}{\sqrt{2}k}\frac{1}{\sqrt{k_{y}^{2}+k_{z}^{2}}}
\begin{pmatrix}
k_{y}^{2}+k_{z}^{2}\\
-k_{x}k_{y}\pm ikk_{z}\\
-k_{x}k_{z}\mp ikk_{y}
\end{pmatrix}\ ,
\end{equation}
with $k=|\bm{k}|=\sqrt{k_{x}^{2}+k_{y}^{2}+k_{z}^{2}}$.
The eigenvectors are obtained by solving the eigenvalue problem for the curl operator in Fourier space.
The parity symmetry in the primordial vector fossil is broken if the modified displacement field includes two helical components unequally. To simplify our analysis, we assume $\bm{\Psi}_{\rm hel}$ to be purely right-handed, e.g., $\bm{\Psi}_{\rm tot}=\bm{\Psi}_{\rm ZA}+\bm{\Psi}_{\mathcal{R}}$.
Note that both parity symmetric and asymmetric simulations should have identical initial density perturbation in Lagrangian space. This is because the initial density perturbation is solely generated by the same $\bm{\Psi}_{\rm ZA}$ \cite{bernardeau+02}, but not by $\bm{\Psi}_{\rm hel}$, as the latter is divergence-free.
The helical displacement only re-maps the density perturbation at one (Eulerian) position to a different position\footnote{The initial particle velocities in both simulations are purely set by the Zel’dovich approximation since any helical component eventually decays with expansion.} and plays a pivotal role in breaking parity symmetry in the initial spin vector field, as we will show later.

We now describe the generation of $\Phi$ and $\Phi_{\mathcal{R}}$.
In Fourier space, the density fluctuation, $\delta(\bm{k})$, is first constructed to follow the input matter power spectrum, $P(k)$,
\begin{equation}
(2\pi)^{3}\delta_{D}(\bm{k}-\bm{k}') P(k)=\langle\delta(\bm{k})\delta^{*}(\bm{k}')\rangle,
\end{equation}
where its normalization amplitude is defined
\begin{equation}\label{eq:sigma}
    \sigma^{2} \equiv \frac{1}{2\pi^{2}}\int k^{2}P(k)dk,
\end{equation}
and set to $\sigma(z_{\rm init})=0.20$.
Note that this corresponds to familiar $\sigma_{8}(z=0)=0.81$, consistent with CMB measurements \cite{planck18}.
We then obtain $\Phi(\bm{k})$ from $\delta(\bm{k})$ using the Poisson equation $-k^{2}\Phi=4\pi G a^{2}\bar{\rho}\delta$.
Similarly, the pseudo-scalar field $\Phi_{\mathcal{R}}$ is obtained to satisfy
% 
% \begin{equation}
%     \langle k^{2}\Phi_{\mathcal{R}}(\bm{k})k'^{2}\Phi_{\mathcal{R}}^{*}(\bm{k}')\rangle = 
%     (2\pi)^{3}\delta_{D}(\bm{k}-\bm{k}') P_{\mathcal{R}}(k),
% \end{equation}
\begin{equation}
    \left\langle \left| \frac{ k^{2}\Phi_{\mathcal{R}}}{4\pi Ga^{2}\bar{\rho}}\right|^{2} \right\rangle= 
    P_{\mathcal{R}}(k),
\end{equation}
where we consider $P_{\mathcal{R}}(k)$ with high-$k$ suppression on $P(k)$ as
\begin{equation}
P_{\mathcal{R}}(k) = P(k)e^{-k^{2}R_{\rm h}^{2}}.
\end{equation}
We set $R_{\rm h}=0.6\mpcph$ in correspondence to the Gaussian radius of a Lagrangian protohalo of $10^{12}\msolph$.
This is to minimize numerical artifacts generated by the helical displacement on scales smaller than the halos we investigate.
The normalization is set to $\sigma_{\mathcal{R}}(z_{\rm init})=1.45$ to produce large enough parity asymmetry to be detected while maintaining the initial matter power spectrum unchanged.
Qualitatively, $\sigma_{\mathcal{R}}$ determines the degree of primordial parity asymmetry, in analogy to non-gaussianity parameter, $f_{\rm NL}$, describing the departure from Gaussianity \cite{eiichiro&spergel01}.
A universe without vectorial-PV corresponds to the case with $\sigma_{\mathcal{R}}=0$.

Independent random number sets are used for generating $\Phi$ and $\Phi_{\mathcal{R}}$ to make them uncorrelated. We identify halos using the Friends-of-Friends algorithm and utilize only halos more massive than $10^{12}\msolph$ at $z=0$. The smallest mass halo contains at least $230$ particles, allowing for a reliable angular momentum measurement \cite{bett+07}. The total number of such halos in the entire 500 simulations of each kind is $N_{\rm halo}\simeq10^{6}$.

\begin{figure}
\includegraphics[clip,width=\columnwidth]{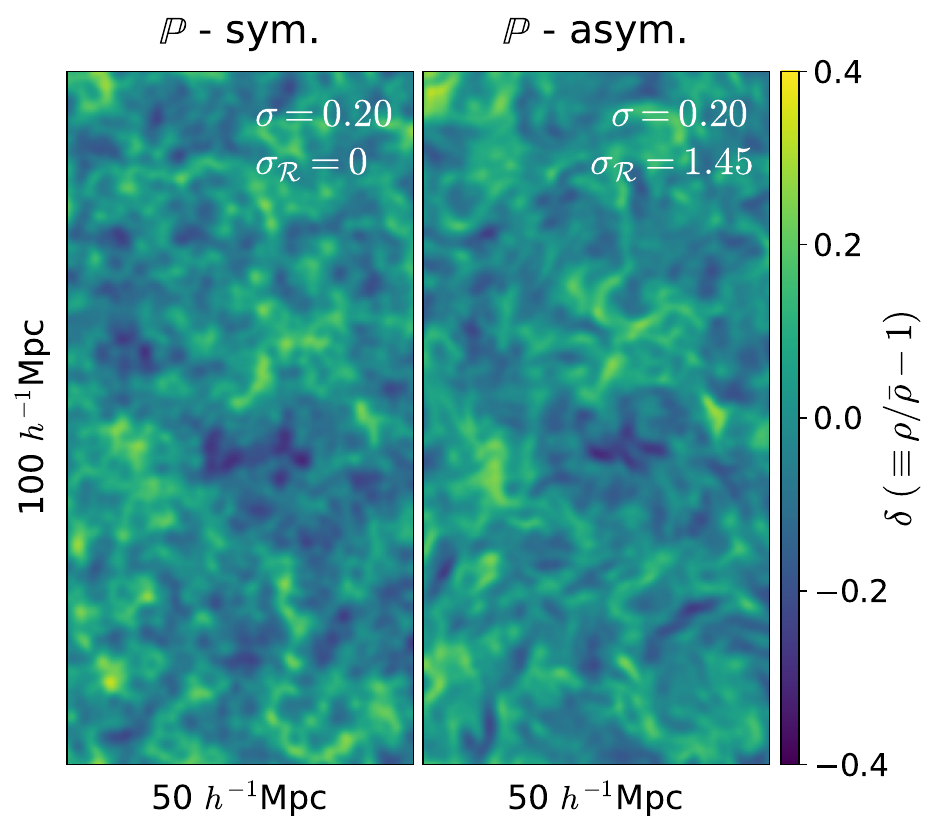}
\caption{(Eulerian) Initial matter density fields at $z_{\rm init}=30$ in the parity symmetric (left) and asymmetric (right) simulations sliced in a $0.4\mpcph$-thickness slab. With the helical displacement (right), matter distribution looks rotationally distorted from that without it (left). Density fields are calculated using the cloud-in-cell (CIC) assignment scheme and smoothed with a Gaussian kernel over $R_{\rm h}=0.6h^{-1}{\rm Mpc}$. 
Note that they look identical in Lagrangian space.}\label{fig:DenF}
\end{figure}

\textit{Initial density field.}---
We begin by visually comparing the initial density fields of the parity symmetric and asymmetric simulations in Fig.~\ref{fig:DenF}.
It can be seen that the helical displacement field $\Psi_{\mathcal{R}}$ contributes to distorting the standard density field similarly to the gravitationally lensed map of the CMB temperature and polarization \cite{zaldarriaga&seljak99,hu&okamoto02,cai+17}.
For example, the density structures in the parity asymmetric realization tend to be stretched and/or warped from their counterparts at similar locations in the standard parity symmetric realization. However, unlike CMB maps exhibiting concentric distortion around over/under-densities, the density field deformation due to the helical displacement occurs incoherently.

\textit{Initial and halo spin fields.}---
We now examine the parity symmetry of the initial conditions. Specifically, we measure the asymmetry between the \RR{}- and \LL{}-helical components of the initial spin vector field.
The initial spin field, $\bm{J}_{\rm init}$, is calculated on $256^3$ grids as a unit vector field constructed from the Hessians of the smoothed initial density, $\rho_{\rm init}^{s}$, and potential field, $\phi_{\rm init}^{s}$, \cite{lee&pen00, codis+15, yu+20, motloch+21, motloch+22},
\begin{equation}\label{eqn:initspin}
     \bm{J}_{\rm init} = \left( J_{\rm init,\alpha} \right) \propto \sum_{\beta\gamma\kappa} \epsilon_{\rm \alpha\beta\gamma} 
     \left( \partial_{\beta}\partial_{\kappa}\rho_{\rm init}^{s}\right)
     \left( \partial_{\gamma}\partial_{\kappa}\Phi_{\rm init}^{s}\right),
\end{equation}
with the Levi-Civita symbol $\epsilon_{\alpha\beta\gamma}$. 
The optimal smoothing scale of the Gaussian kernel is set to $R_{\rm s}=2R_{\rm h}$, suppressing fluctuations on scales smaller than halos \cite{motloch+21,motloch+22,yu+20}.
Note that the initial spin vector field is divergence-free, or purely helical, by construction.
On the other hand, we construct the halo spin field, $\bm{J}_{\rm halo}$, by assigning halo spin vectors, $\bm{j}_{\rm halo}$, to the grids nearest to their respective protohalo centers, $\bm{x}_{\rm ph}$. This is expressed as
\begin{equation}\label{eqn:halospin}
\bm{J}_{\rm halo}(\bm{x}) =
  \begin{cases}
    \bm{j}^{i}_{\rm halo} & \text{if $\bm{x} = \bm{x}_{\rm ph}^{i}$ for $i^{th}$ halo} \\
    \bm{0} & \text{otherwise}.
  \end{cases}
\end{equation}
A halo spin vector represents the unit angular momentum of a halo at $z=0$ and is defined as, 
\begin{equation}
    \bm{j}_{\rm halo}\equiv\frac{\sum_{j}\bm{r}_{p}^{j}\times\bm{v}_{p}^{j}}{|\sum_{j}\bm{r}_{p}^{j}\times\bm{v}_{p}^{j}|},
\end{equation}
where the numerator is the halo angular momentum, calculated as the sum of the angular momenta of its member particles. Here, $\bm{r}_{p}$ and $\bm{v}_{p}$ are particle position and velocity vectors relative to the halo center. 
Similarly, we construct the subsampled initial spin field using only the initial spin vectors that are assigned to the grids nearest to the protohalo centers. Thus, both the halo spin and subsampled initial spin fields have an equal number of nonzero spin vectors at identical positions.

\textit{Measuring helical asymmetry.}---
We examine the parity symmetry of the initial, subsampled initial, and halo spin fields by measuring the differences between their \RR- and \LL-handed components.
We decompose each of the constructed spin fields, $\bm{J}$, into its \RR- and \LL-helical fields,
\begin{equation}
     \bm{J}_{\mathcal{R/L}}=(J_{\mathcal{R/L}})_{\alpha} = \sum_{\beta}
     \mathbb{P}^{\mathcal{R/L}}_{\alpha \beta}({\bm k})J_{\beta}({\bm k}),
\end{equation}
using the projection operator defined
\begin{equation}
     \mathbb{P}^{\mathcal{R/L}}_{\alpha\beta}\equiv\frac{1}{2}
     \left[ (\delta_{\alpha\beta}-\hat{k}_{\alpha}\hat{k}_{\beta})\pm i\sum_{\gamma}\epsilon_{\alpha\beta\gamma}\hat{k}_{\gamma}
     \right].
\end{equation}
The helical asymmetry, $\chi(k)$, of $\bm{J}$ is quantified as 
\begin{equation}\label{eqn:chi}
    \chi(k)\equiv\frac{P_{\mathcal{RR}}(k)-P_{\mathcal{LL}}(k)}{P_{\mathcal{RR}}(k)+P_{\mathcal{LL}}(k)},
\end{equation}
where $P_{{\mathcal{RR}/\mathcal{LL}}}(k)$ represents the power spectrum of the \RR/\LL-helical spin field. It is defined as
\begin{equation}\label{eqn:spinPS}
     \langle \bm{J}_{{\mathcal{R}/\mathcal{L}}}(\bm{k})\cdot\bm{J}^{*}_{{\mathcal{R}/\mathcal{L}}}(\bm {k}') \rangle \equiv (2\pi)^{3}\delta_{D}(\bm{k}-\bm{k}')P_{{\mathcal{RR}/\mathcal{LL}}}(k)\ .
\end{equation}
A spin vector field with perfect helical symmetry satisfies $\chi=0$, whereas $|\chi|$ equals unity when the spin field is maximally asymmetric, or purely \RR/\LL-helical. Note that the cross-power spectrum of \RR{}- and \LL{}-helical fields vanishes because they are orthogonal.

\begin{figure}
\includegraphics[clip,width=0.9\columnwidth]{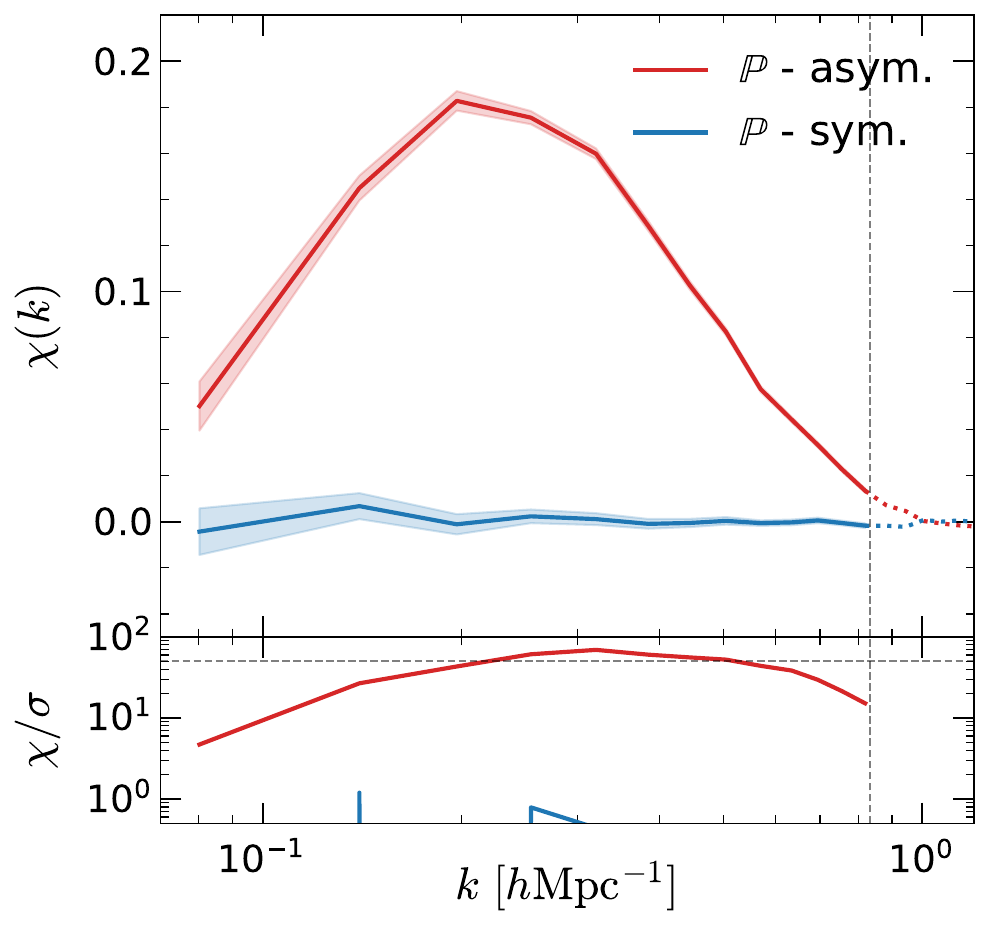}\caption{Average helical asymmetry spectrum (upper) of initial spin fields and the significance of the departure (lower) from the perfect symmetry in the parity symmetric and asymmetric simulations.
The vertical line marks a wavenumber corresponding to twice the Gaussian radius of a $10^{12}\msolph$ protohalo, $k=1/(2R_{\rm h})$, whereas the horizontal line indicates $\chi/\sigma=50$.
Shaded regions represent the standard error of the mean calculated from the 500 realizations, or the total volume of $0.5(h^{-1}{\rm Gpc})^{3}$.
}\label{fig:spinPS}
\end{figure}

In Fig.~\ref{fig:spinPS}, we show the average helical asymmetry in the full initial spin fields for the parity symmetric/asymmetric realizations. We focus only on scales larger than $k=1/(2R_{\rm h})$, below which the helical asymmetry feature in halo spins is likely to be convolved with the halo exclusion effect \cite{baldauf+13}.
In the parity asymmetric realizations, the initial spin field exhibits scale-dependent helical asymmetry that deviates from $\chi=0$, yielding about $70\sigma$ departure at maximum at $k\approx0.3h{\rm Mpc}^{-1}$.
Overall, the initial spin field of the parity asymmetric realization is preferentially \RR-handed, e.g., $\chi>0$. In particular, the helical asymmetry can be as large as $\chi_{\rm peak}\approx0.18$ indicating that the \RR-helical mode is stronger by $44\%$ than the \LL-helical mode.
Such an asymmetry peak becomes higher and shifts toward a larger wavenumber if the initial spin field is constructed adopting a larger smoothing scale.
On the other hand, regardless of the smoothing scale, we observe a nearly perfect helical symmetry (i.e., $\chi\approx0$) on all scales investigated in the parity symmetric simulations.
It is worth noting that the net helicity of the displacement field determines the relative dominance between the two helical modes of the initial spin fields.

\begin{figure}
\includegraphics[clip,width=\columnwidth]{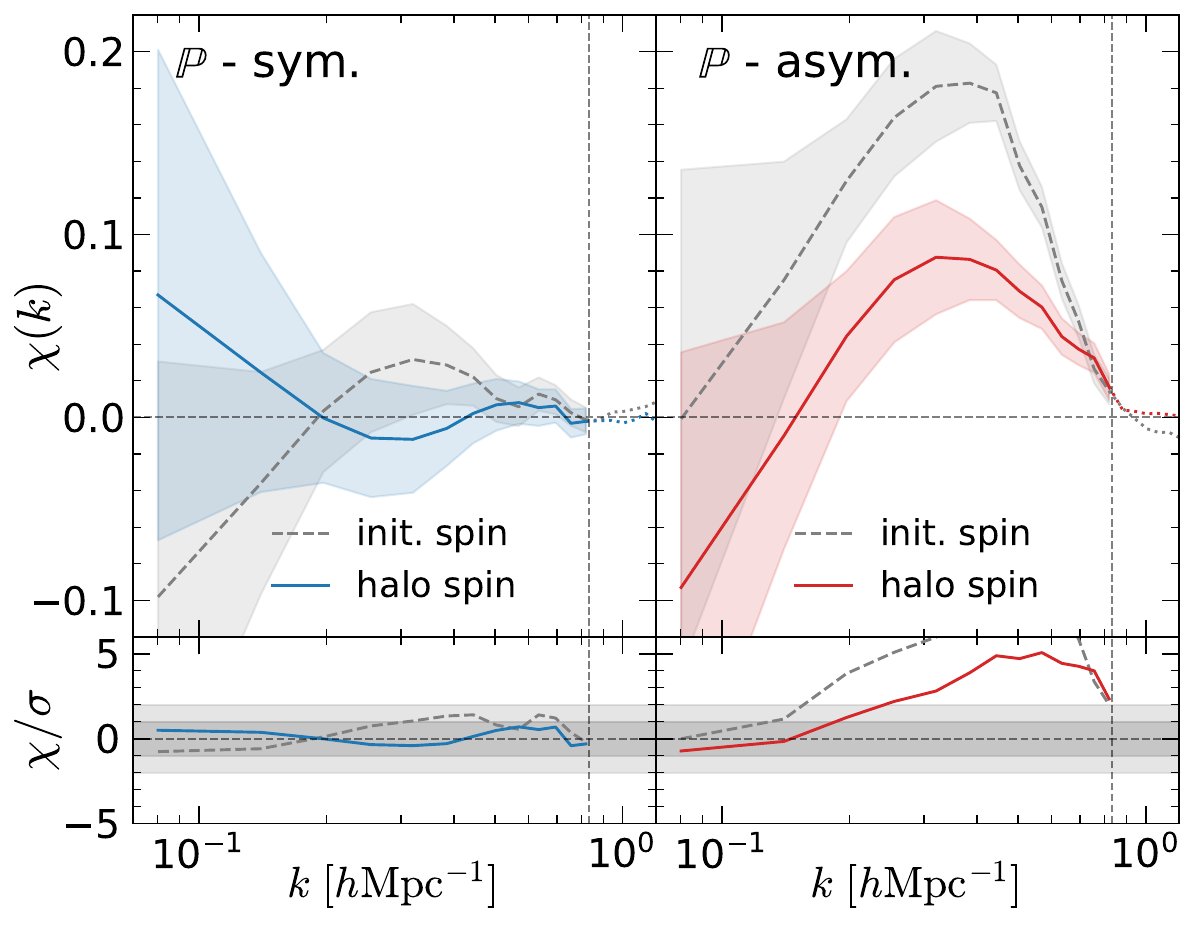}
\caption{Same as Fig.~\ref{fig:spinPS} but for the subsampled initial spin and late-time halo spin fields.
Shaded regions represent the standard errors of the mean using $\sim10^6$ halos. The vertical line again marks $k=1/(2R_{\rm h})$ and grey horizontal bands correspond to $\chi/\sigma\le1$ and $2$.
}\label{fig:DWspinPS_both}
\end{figure}

We now compare the late-time halo spin field with the subsampled initial spin field in Fig.~\ref{fig:DWspinPS_both}.
Unlike the full initial spin field, both spin fields contain shot-noise contamination because a finite number of spin vectors are used for their construction.
To compensate for the shot-noise effect, we rescale by multiplying the same rescaling factor to their helical asymmetry spectra. Consequently, the subsampled and full initial spin fields for the parity asymmetric case have identical peak height, e.g., $\chi_{\rm peak}\simeq0.18$.
In the parity symmetric cases, the subsampled initial field shows larger fluctuations around $\chi=0$ than in the full initial spin field.
Nonetheless, it is still consistent with the helical symmetry within $1.4\sigma$ significance. Similarly, we do not detect any significant deviation from the perfect symmetry in the halo spin field. The departure at maximum is only about $0.7\sigma$-level across the scales investigated. This indicates that the halo spin field remains parity symmetric if the initial spin field were parity symmetric.

On the other hand, in the parity-violating realizations, the subsampled initial spin field shows notably large helical asymmetry, with scale dependence resembling that of the full initial spin field.
The significance of the signal exceeds $6\sigma$ in the range of scale $0.3\le k/[h\,{\rm Mpc}^{-1}] \le0.7$.
For the halo spin field, we again detect considerable helical asymmetry. In particular, the positive departure from perfect symmetry is detected more than $3\sigma$-significance on scales $0.35\le k/[h\,{\rm Mpc}^{-1}] \le0.75$, reaching a maximum of about $5.1\sigma$.
On these scales, the halo spin field preserves on average more than $50\%$ of the initial helical asymmetry.
This clearly demonstrates that the signature of vectorial-PV largely remains even after the gravitational collapse, persisting prominently across these non-linear scales.

\textit{Conclusion and discussion.}---
We assessed the detectability of parity violation in primordial vector fossils using galaxy spins. Utilizing simulations, we examined how effectively the primordial helical asymmetry is preserved in the late-time galaxy spin field.
To this end, we developed a new method to generate initial conditions with parity-violating four-point asymmetry as quantified by the power spectrum of the quadratic initial spin field. Our approach provides a clean channel that effectively embeds parity asymmetry without altering the power spectrum or generating kurtosis via helical remapping, similarly to CMB lensing.
This differentiates our approach from the implementation proposed in \cite{coulton+24}, where larger parity asymmetry inevitably accompanies larger kurtosis and variance.

In the parity asymmetric simulations with $\sigma_{\mathcal{R}}=1.45$, we detected substantial helical asymmetry in the initial spin field, reaching a maximum amplitude of $\chi\simeq0.18$. More than $50\%$ of such asymmetry on a range of scale survived in the late-time halo spin field and was detected maximally at $5.1\sigma$ significance from a total volume of $0.5 h^{-3}{\rm Gpc^{3}}$.
This demonstrates that the imprint of primordial vectorial-PV persists prominently in halo spins, only partly reduced by the non-linear gravitational evolution.
On the other hand, in the null test (i.e., $\sigma_{\mathcal{R}}=0$), the initial spin was helically symmetric, and so was the halo spin field. Our findings validate the robustness of the halo spin field as tracers of the initial spin field, thereby highlighting its efficacy as a tool for probing the vectorial-PV in the early universe.

Such well-preserved parity asymmetry seen in the halo spins is expected to manifest in observable galaxy spins as well. This is because the spin of baryon components of a galaxy well aligns with its initial spin \cite{sheng+23} and host halo spin \cite{hahn+10,deason+11,teklu+15,sheng+23}. Therefore, the initial parity asymmetry should be directly detectable in observation via galaxy spins.
If the observed galaxy spin field is helically asymmetric, it would be an unambiguous indication of the vectorial-PV of primordial origin.

Considering the upcoming Y1-data release of the Dark Energy Spectroscopic Instrument (DESI) \cite{desi}, it presents a timely and compelling opportunity to investigate helical asymmetry in low-redshift galaxy spin fields.
The DESI Bright Galaxy Sample \cite{hahn+23} will provide high-quality photometric data for over 10 million nearby galaxies, offering precise measurements of position angle, axial ratio, color gradient \cite{liao&cooper23}, projected rotational velocity, and winding direction \cite{iye+19}, with which 3-dimensional galaxy spins can be unambiguously determined \cite{motloch+21}.

In light of this, we perform a simple forecast on the detection significance in observation.
To this end, we create mock observed galaxy spin fields from the halo spin fields by reorienting halo spin vectors twice by $\theta$ in random directions. The two reorientations reflect that galaxy spins are typically more misaligned from their initial spins than halo spins \cite{sheng+23} and that spin determination is subject to observational errors.
For our median halo mass, the average angle between a halo and its galaxy spins is $21^{\circ}$ \cite{sheng+23}. Thus, we adopt $\theta=21^{\circ}$ and assume the observational error is identical to $\theta$ for simplicity.
In the mock galaxy spin fields, the helical asymmetry decreases by $18\%$ in $0.35 < k/[h{\rm Mpc}^{-1}] < 0.75$ relative to the halo spin fields.
With such $10^{7}$ galaxy spins, the maximum detection significance for our parity-violating model with $\sigma_{\mathcal{R}}=1.45$ could reach $13\sigma$, assuming that the statistical uncertainty in helical asymmetry measurement is reduced by a factor of $1/\sqrt{10}$, compared to the present analysis using $10^{6}$ spins.

Our approach potentially enables us to place observational constraints on theories involving parity-breaking in vector fossils, with the flexibility to accommodate various $\sigma_{\mathcal{R}}$ and $P_{\mathcal{R}}(k)$. An illustrative example closely related to our study would be the generation of the compensated isocurvature perturbations (CIP) \cite{gordon&lewis03,gordon&pritchard09, grin+11}. In such scenarios, the helical displacement vector field is introduced to displace dark matter relative to the baryon so that their density perturbations have the same amplitude but opposite signs \cite{vanzan+24}. If the CIP displacement field exhibits a preferred handedness, i.e., helical asymmetry, the late-time galaxy spin field can be used for constraining these mechanisms.

Finally, exploring how vectorial-PV manifests in various observables presents another intriguing avenue. As proposed in \cite{vanzan+24}, further refinement of the vectorial-PV with the helical displacement field might potentially elucidate the emergence of parity-odd 4-point correlations observed in galaxy distributions \cite{philcox22, hou+23}. Hence, there is considerable interest in quantifying these 4-point statistics of density fields using our dataset.
Another potential signature of vectorial-PV might appear in the intrinsic alignment of galaxy shapes. While the signature of tensorial-PV is expected to appear in the galaxy shape statistics \cite{biagetti+20, philcox+24shape, okumura&sasaki24}, it is yet unexplored for the vectorial-PV.
Halo shapes are spin-2 tensors that can be decomposed into tensor and vector modes. Therefore, we might be able to detect asymmetry between the two helical states of these modes.
One can also expect to detect a non-zero signal in the inner product between galaxy spin and velocity vectors, as done in \cite{coulton+24}. Because this is a parity-odd quantity, its non-zero measurement using our data will confirm the impact of vectorial-PV.
Such analyses are left for future exploration.

\section{acknowledgments}
We thank anonymous referees for helpful comments that helped improve the original manuscript.
We also thank Donghui Jeong, Eiichiro Komatsu, Oliver Philcox, Shi-Fan Chen, and Tomomi Sunayama for providing valuable feedback on the manuscript. We also thank Academia Sinica Institute of Astronomy and Astrophysics (ASIAA) for hosting the workshop \href{https://events.asiaa.sinica.edu.tw/workshop/20231204/}{`\emph{Large-scale Parity Violation Workshop}'} during which this project was advanced. J.S. acknowledges the support by ASIAA. H.R.Y. is supported by National Science Foundation of China grant No. 12173030. T.O. acknowledges support from the Ministry of Science and Technology of Taiwan under grants No. MOST 111-2112-M- 001-061- and No. NSTC 112-2112-M-001-034- and the Career Development Award, Academia Sinica (AS-CDA-108-M02) for the period of 2019-2023. We thank ASIAA for providing computing resources (High Performance Computing Systems).

% \bibliography{apssamp}% Produces the bibliography via BibTeX.

\input{apssamp.bbl}

\end{document}